\documentstyle[preprint,aps]{revtex}
\begin{document}
\preprint{UM-P-97/4, RCHEP-97/1}
\draft
\title{Dirac magnetic monopole and the discrete symmetries}
\author{A.Yu.Ignatiev\cite{byline1} and G.C.Joshi\cite{byline2}}
\address{Research Centre for High Energy Physics, School of
Physics, University of Melbourne, Parkville, 3052, Victoria,
Australia}
\maketitle
\begin{abstract}
We examine several issues related to the processes of Dirac
monopole-antimonopole production in high-energy collisions
such as $e^+e^-$ annihilation.
Perturbative calculations for such processes are known to be
inherently ambiguous due to the arbitrariness of direction
of the monopole string; this requires use of some prescription
to obtain physical results.
We argue that different prescriptions lead to  drastically
different physical results which suggests that at present we
do not have an entirely satisfactory procedure for the 
elimination of string arbitrariness (this problem is quite
separate from the problems caused by the large coupling
constant). We then analyze the consequencies of discrete
symmetries (P and C) for the monopole production processes and for
the monopole-antimonopole states. The P and C selection rules 
for the monopole-antimonopole states turn out to be different
from those for the ordinary fermion-antifermion
or boson-antiboson system.
In particular, the spin 1/2 monopole and antimonopole should have
{\em the same} helicities if they are produced through the
one-photon annihilation of an electron and positron.
A stronger selection rule holds for {\em spinless} monopoles:
CP symmetry {\em absolutely forbids} the monopole-antimonopole
production through the one-photon annihilation
of an electron and positron. 
Single-photon $e^+e^- \rightarrow g^+g^-$ amplitude has been a key
input in calculating the contribution of virtual $g^+g^-$ pairs to
various physical processes such as the decay $Z\rightarrow 3\gamma$
and the anomalous magnetic moment of the electron. Applying our
conclusions to these cases can lead to significant modifications of
the results obtained in previous works.
  
\end{abstract}

\section{Introduction}
Despite the long history of magnetic monopole, the interplay
between the discrete symmetries and the magnetic charge has not
been completely elucidated in the literature.
Some of the important issues are:
What is the relative parity of the monopole and antimonopole?
Is it negative, as in the case of ordinary fermion-antifermion,
or not?
What happens to the monopole string under the action of C,P,T?
Are there any non-trivial selection rules that are based on discrete 
symmetries and would affect the processes of monopole creation
in particle collisions? One example is the electron-positron
annihilation into monopole-antimonopole pair.
Our paper is an attempt to answer these and related questions.

Motivation, other than theoretical completeness, comes from the
fact that the quantum field theory of magnetic monopole is a
strong coupling theory. Consequently, conventional perturbative
calculations are of little help and one has to rely rather on general
principles such as various symmetries, unitarity and so on.

At present the electromagnetic duality is at the focus of the
elementary particle physics. Various aspects of this concept have been
explored recently leading to significant breakthroughs for instance in
better understanding the structure of the string theory. Another area
of considerable current interest is the study of magnetic monopole and
string solutions in the context of the standard electroweak model.
Also, much interest is generated in investigating the magnetic
monopoles appearing in theories which combine the general relativity
with the Yang-Mills-Higgs systems. All this indicates at the necessity
of a more thorough analysis of the whole complex of problems
associated with the idea of magnetic monopole both from a theoretical
and experimental perspectives (for reviews on Dirac monopoles
\cite{D} see, e.g., \cite{ST,c,rev}). 

Despite several unconfirmed candidate events, the conventional verdict
is that the magnetic monopole has never been observed at the
laboratory. However the experimental efforts devoted to the monopole
searches show no signs of subsiding. Work is being done aimed at
introducing new methods of looking for monopoles, as well as improving
sensitivity of more traditional types of experiments.

There can be different starting points for conducting the magnetic
monopole searches. One can think of producing monopoles directly at
accelerators or observing monopoles bound to the nuclei of the
ferromagnetic materials or, alternatively, look for the bound states
of monopole and antimonopole. In all these cases we speak of
production and detection of monopoles as {\em real} particles
(although they may be confined). Another search strategy is to look
for the effects of {\em virtual} monopoles in high-energy reaction
such as $e^+ e^-$ annihilation and Z-boson decay. Also of interest are
similar effects related to the contribution of virtual magnetic
monopoles to high-precision quantities, for instance, the anomalous
magnetic moment of the electron or muon.

Although all of the experiments for their interpretation have to rely
on the theory governing the behaviour of monopoles, some of the
experiments are in fact less sensitive to the theoretical nuances than
the others. For example, the traditional methods of monopole search            employ only the basic fact that the monopole is the source of the
strong magnetic field around itself. The strength of the magnetic
field is assumed to be dictated by the Dirac quantization condition
\cite{D} that connects the electric and magnetic coupling constants:

\begin{equation}
{eg \over 4\pi}={1 \over 2}n
\end{equation}
Thus, a large value of the magnetic charge and the associated strong
magnetic field is the key signature of the usual experimental methods.

A different class of experiments is based on attempts of producing
monopoles at accelerators or studying the effects of virtual monopoles
in high-energy collisions or in static elementary particle properties.
The latter class of methods for monopole searches attracted 
significant
attention recently (see \cite{amm,r} and references therein).

The main problem with this class of experiment lies in theoretical
interpretation. It is sometimes not recognized clear enough that these
experimental results (usually reported in terms of a specific bound on
the monopole mass) depend much more on the theoretical formulation of
how the monopoles interact than  do ``classical'' search methods. The
difficult question is how to obtain {\em unambiguous} predictions from
the monopole theory. There are two sources of difficulties. The first
is a well-known fact that the coupling constanst (that is, the
magnetic charge of the pole) should be very large if the Dirac
quantization condition is to be true. That makes impossible the use of
perturbation theory for practical calculations (although it
can be used within an effective field theory approach). 

The second difficulty
(which has not been as much popularized) is, perhaps, more
fundamental. It has nothing to do with the magnitude of the coupling
constant at all; rather, it is related to the existence of Dirac
string -- infinitely thin line of magnetic flux stretching from the
magnetic pole to infinity. It has been often repeated in the
literature that the Dirac quantization condition makes the string
invisible. However, in reality the situation is far from being so
simple and clear. This is especially true in the context of quantum
field theory where monopoles are allowed to be created and
annihilated (recall that the Dirac quantization condition was
initially derived for a simple {\em quantum mechanical} system ``
electron plus monopole''). It is generally believed that the full
quantum field theory does not depend on how we choose the position of
the string which can be arbitrary. However, the peculiarity of the
monopole theory is that the formulation of the theory cannot be made
without recourse to the string concept in one or another form. In
other words, the quantum theory of monopoles is not {\em manifestly}
string-independent. Since the string fixes a specific direction, the
theory is not {\em manifestly} Lorentz invariant either. Perhaps, it
is a unique example of a physical theory possessing implicit Lorentz
invariance which nevertheless cannot be formulated in a {\em
manifestly} invariant way.

What are the practical implications of this fundamental theoretical
feature? One consequence is this. Imagine that we forget for a moment
about the large coupling constant and attempt to calculate some
physical quantity in the first order of perturbation theory (such as
the monopole-antimonopole production in $e^+ e^- $ annihilation). The
result will be discouraging because it will be ambiguous. More
exactly, the result will depend explicitely on the string direction            which is clearly unacceptable. Obviously, a serious question is how to
deal with this type of situation.

Consider, for example, the process of $e^+e^-$ annihilation
into monopole-antimonopole pair (assumed to be fermions). It has a
virtue of being physically interesting and simple enough at the same
time. This process has been previously considered and a prescription
has been given for elimination of string dependence \cite{d} which 
has
been subsequently adopted in \cite{amm,r}. The resulting
cross-section is not very
different from the cross-section for the creation of a pair of usual  
fermion-antifermion.

However, we believe that the prescription is not enirely
satisfactory. One reason for concern is that it only
gives the value of 
the {\em squared modulus of the amplitude}, but not {\em the
 amplitude} itself.

Therefore, it would be difficult to generalize it for the cases
when an {\em interference} of two amplitudes is involved (for
instance, if we want to calculate the interference
between electromagnetic and Z-boson contributions to the
monopole-antimonopole production in $e^+e^-$ annihilation).

One purpose of this paper is to consider an alternative
 procedure and see if the physical results would be the same.  More
specifically, we  propose an alternative prescription 
based on the averaging of the amplitude over all possible directions 
of the
string. This procedure has a clear physical meaning since the string
is supposed to be unobservable. However, it leads to a drastically
different answer: according to this prescription, the amplitude
of $e^+e^-$ annihilation into the monopole-antimonopole pair
should be {\em zero} to the lowest order of perturbation theory.

This result suggests that the task of extracting the physically
meaningful results from the inherently ambiguous perturbative
calculations should be considered as an open problem requiring
further investigation.

In this paper we try to circumvent this problem by using only general
 principles
of quantum field theory whose validity does not rely on the use
of perturbation theory. It is natural to start with the consideration
of the role of the discrete symmetries such as C, P and T 
transformations and to see what constraints are provided
by these symmetries.

We show that the behaviour of the monopole-antimonopole
system under discrete symmetries
is rather different from that of standard fermion-antifermion
or boson-antiboson system (standard means
not carrying magnetic charge). In particular, there arise 
selection rules for the process of the monopole-antimonopole
production through one-photon annihilation of an electron
and positron. For spin 1/2 monopole the P and C symmetries 
require that the monopole
and antimonopole have {\em the same} helicities. 
For {\em spinless} monopoles
CP symmetry {\em absolutely forbids} the monopole-antimonopole
production through the one-photon annihilation
of an electron and positron.

The plan of the work is as follows.
 Section 2 summarizes the Feynman rules for the
monopole field theory and the structure of the amplitude $e^+ e^-$
annihilation into monopole-antimonopole pair. The appearance of string
dependence is emphasized. 
In Section 3 we suggest an alternative prescription for eliminating
the string dependence of the amplitude. The averaging of the 
amplitude over the string directions is carried out which results
in vanishing of the one-photon exchange amplitude for the monopole-
antimonopole production in $e^+e^-$ annihilation. 
Section 4 is devoted to the analysis of the discrete symmetries
in the quantum field theory of magnetic monopoles.
The results are applied to the derivation of selection
rules for the monopole-antimonopole system  in Section 5.
 Finally, we present our conclusions in Section 6.

\section{The Feynman rules and the electron-positron annihilation 
into 
monopole-antimonopole}
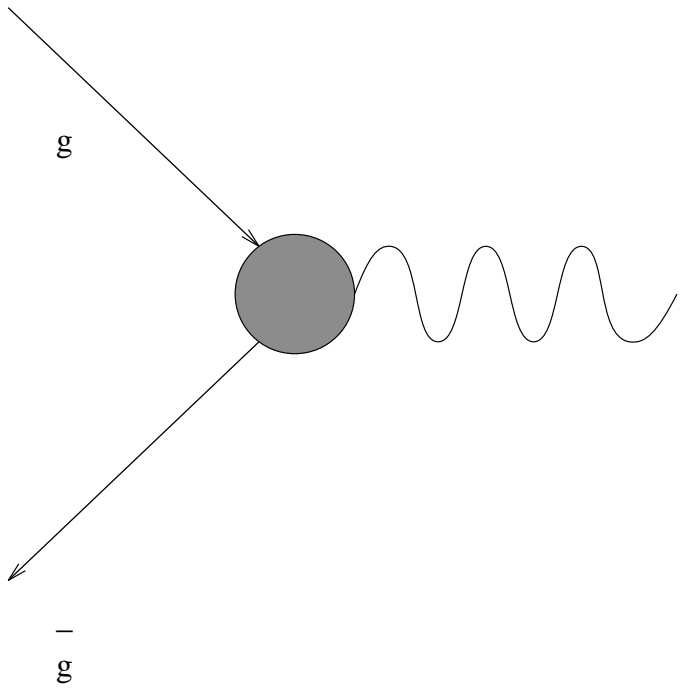
\begin{figure}
\caption{Feynman rules for the photon-monopole interaction}
\label{fig1}
\end{figure}

The Feynman rules \cite{d} describing the interactions of photons 
and monopoles
have the following form (Fig.~\ref{fig1}):
\begin{equation}
-ig{\epsilon^{\mu\nu\lambda\rho}\gamma^{\nu}n^{\lambda}q^{\rho}
\over qn+i \epsilon}.
\end{equation}
The photon and fermion propagators, as well as the photon-electron 
vertex,
remain the same as in the standard QED.
Note that in other formulations of the monopole quantum field theory
the Feynman rules would be different (for details, see \cite{rev}).
The most notable feature of these Feynman rules is the fact that they
depend on the vector $n$ which corresponds to the direction of the
string. Thus, these Feynman rules are not manifestly invariant.
However, it is believed that the full theory is nevertheless Lorentz-
invariant, that is physical predictions should not depend on the
specific direction of the vector $n$.

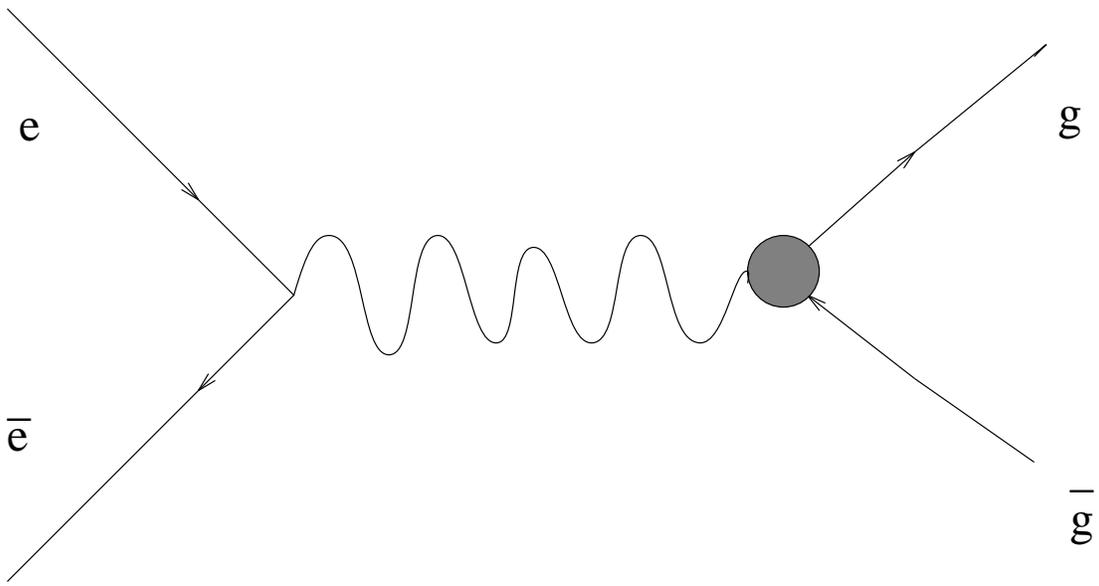
\begin{figure}
\caption{Electron-positron annihilation into monopole-antimonopole}
\label{fig2}
\end{figure}

Now, let us write down the amplitude of the process of the electron-
positron annihilation into the monopole-antimonopole pair. The
amplitude has the following form (Fig.~\ref{fig2}):
\begin{equation}
A=ieg K^{\beta}                                 
\epsilon^{\mu\beta\gamma\delta}
{n^{\gamma}q^{\delta} \over qn}{1 \over q^2}
J^{\mu}.
\end{equation}
where
\begin{equation}
J^{\mu}={\bar v}_e(p_2)\gamma^{\mu}u_e(p_1),
K^{\beta}={\bar u}_g(p_3)\gamma^{\beta}v_g(p_4).
\end{equation}

The dependence on $n$ remains even after the squaring of the amplitude
is made. An obvious question is how to make sense out of the n-
dependent quantity. It has been suggested in Ref.~\cite{d} that one 
should
drop the terms which have no pole in $q^2$ and thus to arrive at the
following result:
\begin{equation}
\label{5}
|A|^2={e^2g^2 \over q^4}[(KJ^{\dag})(JK^{\dag})-
(JJ^{\dag})(KK^{\dag})].
\end{equation}

\section{A different prescription}
However, the consistency of such a prescription can be questioned
on the grounds that it gives the corrected value of the 
{\em squared matrix element} but not of the {\em amplitude} itself.
Therefore, it would be difficult to generalize it for the cases
when an {\em interference} of two amplitudes is involved (for
instance, if we want to calculate the interference
between electromagnetic and Z-boson contributions to the
monopole-antimonopole production in $e^+e^-$ annihilation).
Another concern is whether Eq.~ (\ref{5}) is positively
definite or not.
There exist a different approach to the problem of dealing with
the $n$ dependence. The idea is to average over all possible
directions of $n$. Since there are no physically preferred directions
of $n$, all the directions should be taken with the same weight.
Because all these directions are physically indistinguishable, we have
to perform averaging of the amplitude rather than of the squared
 matrix
element. Therefore, we need  to find the average value:
\begin{equation}
\langle {n^{\gamma} \over qn} \rangle.
\end{equation}
By Lorentz invariance, it is sufficient to find this average value
in a system where $n^0=0$ and, consequently, ${\bf n}^2=1$:

\begin{equation}
\langle {{\bf n} \over - {\bf q}{\bf n}}\rangle =
{1 \over 4\pi} \int {{\bf n} \over - {\bf q}{\bf n}}d\Omega.
\end{equation}
In evaluating this integral one should be careful about a possible
singularity arising when the vector ${\bf n}$ becomes orthogonal
to ${\bf q}$. Let us choose the $z$ axis of the spherical coordinate
system such as to be parallel to ${\bf q}$, and calculate the $x,y,z$
components of the average:
\begin{equation}
{1 \over 4\pi}\int {n_x \over - {\bf q}{\bf n}} d\Omega =
-{1 \over 4\pi|{\bf q}|}\int^{1}_{-1} {\sqrt{1-t^2} \over t}dt
\int^{2\pi}_{0}\cos\phi d\phi ,
\end{equation}
where $t=\cos\theta$. Although the integral over $\phi$ vanishes,
we need to prove that the integral over $t$ is not singular.
For this purpose we have to invoke the $ qn+i\epsilon$ rule
(or, in 3-dimensional terms, the $ {\bf q}{\bf n} -i\epsilon$ rule):
\begin{equation}
\int^{1}_{-1} {\sqrt{1-t^2} \over t}dt \rightarrow
\int^{1}_{-1} {\sqrt{1-t^2} \over t -i\epsilon}dt =
\wp \int^{1}_{-1} {\sqrt{1-t^2} \over t}dt +
i\pi \int^{1}_{-1} \delta (t) \sqrt{1-t^2}dt =i\pi
\end{equation}
Thus, indeed, the $t$-integral is finite and, therefore,
\begin{equation}
{1 \over 4\pi}\int {n_x \over - {\bf q}{\bf n}} d\Omega =0.
\end{equation}
Furthermore, a similar argument shows that the $y$-component
of the average  value also vanishes: 
\begin{equation}
{1 \over 4\pi}\int {n_y \over - {\bf q}{\bf n}} d\Omega =0.
\end{equation}
Now, the $z$-component is;
\begin{equation}
{1 \over 4\pi}\int {n_z \over - {\bf q}{\bf n}} d\Omega =
-{1 \over 4\pi} {1 \over |{\bf q}|} \int d\Omega =
-{1 \over |{\bf q}|}.
\end{equation}
Thus, finally, we obtain:
\begin{equation}
{1 \over 4\pi}\int {{\bf n} \over - {\bf q}{\bf n}} d\Omega =
- {{\bf q} \over {\bf q}^2}.
\end{equation}
Consequently,
\begin{equation}
\langle {n^{\gamma} \over qn} \rangle={q^{\gamma} \over q^2}.
\end{equation}
Now, if we insert this value into the $e^+e^-$ annihilation
amplitude, we obtain
\begin{equation}
A=ieg K^{\beta}                                 
\epsilon^{\mu\beta\gamma\delta}
{q^{\gamma}q^{\delta} \over q^4}
J^{\mu}=0.
\end{equation}
Thus, we arrive  to the same conclusion: {\em  if one uses the
averaging procedure to eliminate the string dependence of the
 amplitude,
than  the one-photon
amplitude of the $e^+e^-$ annihilation into monopole-antimonopole pair
turns out to be zero}.
To summarize, we have shown that two different prescriptions used
to eliminate the string dependence of the amplitude lead to 
drastically different physical results. Therefore, we suggest
to try to circumvent this problem by using only general principles
of quantum field theory whose validity does not rely on the use
of perturbation theory. It is natural to start with the consideration
of the role of the discrete symmetries such as C, P and T 
transformations and to see what constraints are provided
by these symmetries.  

\section{Discrete symmetries in the quantum field theory of magnetic
monopoles}
There exist several formulations of the quantum field theory with
electric and magnetic charges. However, all of the formulations have
been shown \cite{rev}
to be equivalent (except the formulation due to Cabibbo and
Ferrari). Therefore, we will not need to specify exactly in which
theoretical context are going to work. Rather, we will focus on the
properties of the quantum field theory under the action of the
discrete symmetries such as space reflection and charge conjugation.
It can be shown \cite{ram,w,z}
that the quantum field theory of the electric and
magnetic charges is invariant under the following discrete
transformations (we use Majorana representation, and denote the
magnetically charged fields by the subscript $g$):
\begin{equation}
C:\;\;{\bf E},{\bf H},\psi,\psi_g \rightarrow -{\bf E},-{\bf H},
\psi^{\dag},\psi_g^{\dag}.
\end{equation}
\begin{equation}
\label{17}
P:\;\;{\bf E(x)},{\bf H(x)},\psi{\bf (x)},\psi_g{\bf (x)} \rightarrow
-{\bf E(-x)},-{\bf H(-x)},\gamma^0\psi{\bf (-x)},\gamma^0
\psi_g^{\dag}{\bf (-x)}
\end{equation}
\begin{equation}
T:\;\;{\bf E}(t),{\bf H}(t),\psi(t),\psi_g(t) \rightarrow {\bf E}(-
t),-{\bf H}(-t),\gamma^0 \gamma^5
\psi (-t),\gamma^0 \gamma^5\psi_g^{\dag}(-t).
\end{equation}
Here, a comment on terminology is in order. There is some confusion in
the literature as to whether we should retain the names ``P
reflection'' and ``T inversion'' for the above operations or we should
call them ``PM'' and ``TM'' transformations, where M stands for the
inversion of the magnetic charge. However, this difference is of
semantical rather than of physical character;  switching from one
terminology to the other does not entail any physical consequences. In
this paper we adopt the the first point of view, i.e. we keep the
names parity and T inversion for the operations we have just
introduced without making any further qualifications (the same view
is adopted in \cite{w}).

Sometimes one can find in the literature the statements to the effect
that the theory of monopoles is not invariant under P and T
symmetries. These statements refer to the situation when the discrete
symmetries are assumed to act on the magnetically charged particles in
exactly the same way as they act on the electrically charged
particles, that is their action on the magnetically charged states
does not include the sign inversion of the magnetic charge. It is easy
to see that if the discrete transformations are defined in that way,
then the theory is indeed P and T non-invariant.  However, the
possibility to define P and T symmetries in such a way that they are
conserved makes the ``non-conserving'' definition irrelevant.

Note also that we assume that there are no particles carrying
simultaneously both the electric and magnetic charge; in other words,
there are no dyons in the theory; in this case the conserving P and T
operations do not exist \cite{z}.

Now, we need to write these transformations in terms of creation
(or annihilation ) operators rather than in terms of local fields.
Let us first recall the standard formulas for spinor fields in the
Majorana representation (we follow the Bjorken-Drell notation 
\cite{BD}):
\begin{eqnarray}
\psi ({\bf x},t)= &&{1 \over (2\pi)^{3/2}} \int d^3p \sum_{s}
\sqrt{{m \over E}}
[a(p,s)u(p,s) \exp (-iEt + i{\bf px})\nonumber\\
&&+b^{\dag}(p,s)v(p,s) \exp (iEt - i{\bf px}).
\end{eqnarray}
We are working in the Majorana representation which is connected with 
the standard one (i.e., with diagonal $\gamma^0$) via the following
relationships: 
\begin{eqnarray}
\gamma^{\mu} & = & U\gamma^{\mu}_S U^{\dag}\\
u(p,s) & = & Uu_S(p,s)\\
v(p,s) & = & Uv_S(p,s),
\end{eqnarray}
where the subscript $S$ marks the standard representation and the
transition matrix $U$ is 
\begin{equation}
 U = \left( \begin{array}{cc}
I & \sigma_2 \\
\sigma_2 & -I
\end{array}
\right).
\end{equation}
The spinors $u_S$ and $v_S$ are defined according to:
\begin{equation}
u_S(p,s)=\sqrt{{E+m \over 2m}}
\left( \begin{array}{c}
\chi_1({\bf s}_0) \\
{{\bf \sigma p} \over E+m}\chi_1({\bf s}_0)
\end{array}
\right)
\end{equation}
\begin{equation}
v_S(p,s)=\sqrt{{E+m \over 2m}}
\left( \begin{array}{c}
{{\bf \sigma p} \over E+m}\chi_2({\bf s}_0)\\
\chi_2({\bf s}_0) 
\end{array}
\right).
\end{equation}

Here, ${\bf s}_0$ is the spatial part of the spin 4-vector $s_{\mu}$
taken in the rest system of the 4-vector $p_{\mu}$; the 2-column
spinors $\chi_1({\bf s}_0)$ and $\chi_2({\bf s}_0)$ correspond to
the spin parallel and antiparallel to the direction ${\bf s}_0$:
\begin{eqnarray}
{\bf \sigma s_0}\chi_1({\bf s}_0) &=& \chi_1({\bf s}_0)\\
{\bf \sigma s_0}\chi_2({\bf s}_0) &=& -\chi_2({\bf s}_0).
\end{eqnarray}
Thus, explicitely, we have in the Majorana representation:
\begin{equation}
\gamma^0= \left( \begin{array}{cc}
0 & \sigma_2 \\
\sigma_2 & 0
\end{array}
\right) \;\;\;
\gamma^1=\left( \begin{array}{cc}
i\sigma_3 & 0 \\
0 & i\sigma_3 
\end{array}
\right)
\end{equation}
\begin{equation}
\gamma^2= \left( \begin{array}{cc}
0 & -\sigma_2 \\
\sigma_2 & 0
\end{array}
\right) \;\;\;
\gamma^3=\left( \begin{array}{cc}
-i\sigma_1 & 0 \\
0 & -i\sigma_1 
\end{array}
\right).
\end{equation}
All the $\gamma$-matrices are purely imaginary which is the 
characteristic of the Majorana representation.
Therefore, the spinors $u(p,s)$ and $v(p,s)$ are:
\begin{equation}
\label{30}
u(p,s)= Uu_S(p,s)=
\sqrt{{E+m \over 2m}}
\left( \begin{array}{c}
(1+{\sigma_2{\bf \sigma p} \over E+m} )\chi_1({\bf s}_0)\\
(\sigma_2 - {{\bf \sigma p} \over E+m})\chi_1({\bf s}_0)
\end{array}
\right).
\end{equation}
\begin{equation}
v(p,s)= Uv_S(p,s)=
\sqrt{{E+m \over 2m}}
\left( \begin{array}{c}
(\sigma_2+{{\bf \sigma p} \over E+m} )\chi_2({\bf s}_0)\\
(-1 +{\sigma_2{\bf \sigma p} \over E+m})\chi_2({\bf s}_0)
\end{array}
\right).
\end{equation}
The relative phase of the spinors $u$ and $v$ is chosen in such a way
that the following equality holds:
\begin{equation}
\label{31}
u^*(p,s)=v(p,s).
\end{equation}

Now, the action of the parity operator on the magnetically charged
field $\psi_g$ reads:
\begin{eqnarray}
P\psi_g ({\bf x},t)P^{-1}=&& {1 \over (2\pi)^{3/2}} \int d^3p \sum_{s}
\sqrt{{m \over E}}
[Pa_g(p,s)P^{-1}u(p,s) \exp (-iEt + i{\bf px})\nonumber\\
&&+Pb_{g}^{\dag}(p,s)P^{-1}v(p,s) \exp (iEt - i{\bf px}).
\end{eqnarray}
On the other hand, using the rule (\ref{17}) we can write:
\begin{eqnarray}
P\psi_g ({\bf x},t)P^{-1}=&&{1 \over (2\pi)^{3/2}} \int d^3p \sum_{s}
a_{g}^{\dag}(p,s)\gamma^0u^*(p,s) \exp (iEt + i{\bf px})\nonumber\\
&&+b_{g}(p,s)\gamma^0v^*(p,s) \exp (-iEt - i{\bf px}).
\end{eqnarray}

Now, using Eq.~(\ref{30}) and (\ref{31}) one can show that
\begin{equation}
\gamma^0u^*(p,s)=-v(-p,s), \;\;\; \gamma^0v^*(p,s)=u(-p,s).
\end{equation}
Therefore, we obtain the following law of transformation of the
creation and annihilation operators of a magnetically charged 
fermion:
\begin{equation}
Pa_g(p,s)P^{-1}=b_g(-p,s),\;\;\; Pb_{g}^{\dag}(p,s)P^{-1}=
-a_{g}^{\dag}(-p,s).
\end{equation}
For a magnetically uncharged fermion $\psi$, the transformation law
is:
\begin{equation}
Pa(p,s)P^{-1}=a(-p,s), \;\;\; Pb(p,s)P^{-1}=-b(-p,s).
\end{equation}
In a similar fashion we can derive the laws of C transformation of
a magnetically charged fermion:
\begin{equation}
Ca_g(p,s)C^{-1}=b_g(p,s),  \;\;\; Cb_{g}(p,s)C^{-1}=
a_{g}(p,s).
\end{equation}
For a fermion without magnetic charge, the C transformation has the
 same form.
In a similar way one can obtain the formulas for the T reversal
but we will not need to use them in the present paper.
Hence, we see a clear difference between
the behavior of the states with the electric charge and the
magnetically charged states. The parity and time inversion acting on
the electrically charged states do not change the electric charge of
these states, that is under P transformation the electron is carried
into an electron with opposite momentum and, likewise, positron is
transformed into positron state with the opposite momentum. Similarly,
under time inversion the electron state is transformed into the
electron state with opposite momentum and spin; the positron is turned
into the positron with opposite momentum and spin. So, the P and T
transformation do not change the electric charge at all. On the
contrary, for magnetically charged particles the situation is exactly
opposite: the P and T reflections necessarily include {\em the 
change of
sign} of the magnetic charge. For instance, P transformation acting on
the magnetic monopole
takes it into antimonopole with the opposite momentum; likewise, under
P parity the antimonopole is transformed into monopole with the
opposite momentum. The same is true for T reversal: the T
transformation changes the monopole into antimonopole with opposite
momentum and spin; the antimonopole is changed into monopole with
inverse momentum and spin.

\section{Discrete symmetries and the monopole string}
So far we have completely ignored the existence of a string
(that is, the infinitely thin line of infinitely strong
magnetic field) attached to the magnetic monopole. Note that it
is possible to formulate {\em quantum mechanics} of the 
magnetic charge in such a way as to avoid introduction of
the string \cite{wy}. However, at the level of {\em quantum
field theory} all known formulations do introduce the string 
under different guises \cite{rev}. It is therefore of obvious
importance to know how the discrete symmetries act on the string,
if at all.

We would like to stress that our considerations in this 
section are of heuristic rather than of rigourous  character.
We try to pinpoint those aspects of the problem that would be
common to all specific theories of quantum electromagnetodynamics,
rather than making a theory-by-theory analysis.

We should, however, make one very important distinction from
the beginning. Namely, we have to distinguish between two types
of strings: first, the semi-infinite string and, second, the string
that is infinite in both directions. For brevity, we shall call
them ``short strings'' and ``long strings'', respectively.
Let us consider the short string first. Denote by ${\bf n}$ 
the unit vector in the direction of the string and by ${\bf H}$
the magnetic field of the string.
Under charge conjugation (which, by definition, includes both
electric and magnetic charge reversal) we have (see Fig.3a):
\begin{equation}
{\bf H} \rightarrow -{\bf H} \;\;\; g \rightarrow -g \;\;\;
{\bf n} \rightarrow {\bf n}.
\end{equation}
Next, the parity transformation P acts as follows (Fig.3b):
\begin{equation}
{\bf H(x)} \rightarrow {\bf H(-x)} \;\;\; g \rightarrow -g \;\;\;
{\bf n} \rightarrow -{\bf n}.
\end{equation}
Performing the time reversal T, we obtain (Fig.3a):
\begin{equation}
{\bf H}(t) \rightarrow -{\bf H}(-t) \;\;\; g \rightarrow -g \;\;\;
{\bf n} \rightarrow {\bf n}.
\end{equation}
Under CP parity the transformation law is (Fig.3c):
\begin{equation}
{\bf H(x)} \rightarrow -{\bf H(-x)} \;\;\; g \rightarrow g \;\;\;
{\bf n} \rightarrow -{\bf n}.
\end{equation}
Finally, under CPT we have (Fig.3b):
\begin{equation}
{\bf H}({\bf x},t) \rightarrow {\bf H}(-{\bf x}, -t) \;\;\;
 g \rightarrow -g \;\;\;
{\bf n} \rightarrow -{\bf n}.
\end{equation}
For the case of the long string (i.e., infinite in both directions)
the discrete transformations look as follows (Fig.4): 
\begin{equation}
C: \;\;\; {\bf H} \rightarrow -{\bf H} \;\;\; g \rightarrow -g 
\end{equation}
\begin{equation}
P: \;\;\; {\bf H(x)} \rightarrow -{\bf H(-x)} \;\;\; g \rightarrow -g 
\end{equation}
\begin{equation}
T: \;\;\; {\bf H(t)} \rightarrow -{\bf H(-t)} \;\;\; g \rightarrow -g 
\end{equation}
\begin{equation}
CP: \;\;\; {\bf H(x)} \rightarrow -{\bf H(-x)} \;\;\; g \rightarrow g.
\end{equation}
\begin{equation}
CPT: \;\;\; {\bf H}({\bf x},t) \rightarrow {\bf H}({\bf -x},-t) 
\;\;\; g \rightarrow -g. 
\end{equation}
Thus we are led to the following conjecture about the behavior
with respect to discrete symmetries of the quantum field theories
describing magnetic monopoles with short (i.e. semi-infinite)
strings.  These theories either conserve CPT,
 P and CP symmetries or they violate all of them,  CPT, P 
and CP, simultaneously.
Thus, according to this conjecture it would be hard to conceive
a CPT invariant magnetic monopole theory that would violate
parity. Another interesting question is whether one can construct
an example of CPT non-invariant theory along these lines.
Of course, that would not contradict the famous CPT theorem
because one of the requirements for this theorem to be true is
the condition of locality whereas the monopole string appears
to be a non-local object.

\section{Discrete symmetries and physical processes}

Now we are in a position to apply the discrete symmetries
to consideration of specific physical processes in order
to establish whether any selection rules can be obtained 
or not.
Since the monopoles are expected to be relativistic, let us 
use the helicity basis for their consideration. In this
basis the pair of monopole-antimonopole is described by
a wave function $\psi_{JM\lambda_1\lambda_2}$ where $J$ is
the total angular momentum of the pair, $M$ is the projection
of $J$ and $\lambda_1$ and   $\lambda_2$  are the helicities of 
the monopole and antimonopole.
The action of discrete symmetries is given by:
\begin{equation}
P\psi_{JM\lambda_1\lambda_2}=\psi_{JM-\lambda_2 -\lambda_1},
\end{equation}
\begin{equation}
C\psi_{JM\lambda_1 \lambda_2}=(-1)^J\psi_{JM\lambda_2 \lambda_1}.
\end{equation}
Using these rules, we can construct the wave function that has 
the photon quantum numbers, i.e. $J=1$, $P=-1$ and $C=-1$:
\begin{equation}
\label{wf}
\psi_{1M}={1\over \sqrt{2}}(\psi_{1M {1 \over 2}{1 \over 2}}
            -\psi_{1M -{1 \over 2}-{1 \over 2}}).
\end{equation}
Thus we see that in order to couple to the photon, the monopole
and antimonopole should have the {\em same} helicities.
To further understand the physical meaning of this condition,
let us consider the non-relativistic limit, in which the monopole
-antimonopole pair is described by the wave function 
$\psi_{JLSM}$ where $L$ and $S$ are the total orbital momentum
and spin, respectively. The connection between the wave functions
$\psi_{JLSM}$ and $\psi_{JM\lambda_1\lambda_2}$ is given by 
\cite{BLP}:
\begin{equation}
\psi_{JLSM}=\sum_{\lambda_1\lambda_2}
\psi_{JM\lambda_1\lambda_2}
\langle JM \lambda_1\lambda_2 | JLSM \rangle,
\end{equation}
where the coefficients are expressed through the $3j$ symbols
as follows:
\begin{equation}
\langle JM \lambda_1\lambda_2 | JLSM \rangle=
(-i)^L(-1)^S \sqrt{(2L+1)(2S+1)}
\left( \begin{array}{ccc}
{1 \over 2}&{1 \over 2}&S \\
\lambda_1& -\lambda_2&- \Lambda 
\end{array} \right)
\left( \begin{array}{ccc}
L&S&J\\
0& \Lambda &-\Lambda 
\end{array} \right).
\end{equation}

It can be shown that the wave function Eq.~(\ref{wf}) corresponds
to the state $S=0$, $L=1$ in the non-relativistic limit, i.e.:
\begin{equation}
\psi_{110M} = {1\over \sqrt{2}}(\psi_{1M {1 \over 2}{1 \over 2}}
            -\psi_{1M -{1 \over 2}-{1 \over 2}}),
\end{equation}
\begin{equation}
\Lambda=\lambda_1-\lambda_2
\end{equation}
In other words, the $J^{PC}=1^{--}$ of the monopole-antimonopole
pair corresponds to the $^1P_1$ state in the non-relativistic 
limit. This should be contrasted with the case of the standard
fermion-antifermion pair (such as positronium or quarkonium)
for which the $1^{--}$ state is $^3S_1$ (or $^3D_1$). 

Now, let us consider spin 0 monopoles for we do not have any 
evidence concerning the possible value of the monopole spin.
From the similar considerations as the above, it can be shown
that the spinless monopole-antimonopole system has the following
quantum numbers:
\begin{equation}
\label{s}
P=1, \;\;\; C=(-1)^J,
\end{equation}
where $J$ is the total angular momentum of the system.
Thus, the spinless monopole-antimonopole production through the
one-photon $e^{+}e^{-}$ annihilation is {\em absolutely
forbidden}.  
Next, it follows from Eq.~(\ref{s})
that in the state with the total angular momentum
$J=1$ the monopole-antimonopole pair has always $CP=-1$.
Therefore, CP symmetry absolutely forbids the $1^{--}$
and $1^{++}$ states of the monopole-antimonopole system.
Note that this conclusion holds true even if P and C parities
do not conserve separately, but CP does.
This means that the the decay of Z-boson into  spin 0
 monopole-antimonopole pair would be absolutely forbidden in 
a CP invariant theory. 

Thus we have shown that C and P invariance imposes exact
selection rules on the monopole-antimonopole state produced
through the one-photon channel of $e^+e^-$ annihilation.

It remains to be investigated whether these selection rules
can help us to understand why the monopoles have not been
observed experimentally. 

Recently the contribution of virtual monopoles to various physical
processes has been examined in several papers. One of them was the
contribution of virtual monopole-antimonopole pairs to the anomalous
magnetic moment of the electron \cite{amm} (see Fig. 5).
Another process is the monopole loop
contribution to the decay of Z boson into 3 photons \cite{r}.
Single-photon $e^+e^- \rightarrow g^+g^-$ amplitude has been a key
input in calculating the contribution of virtual $g^+g^-$ pairs to
these processes. Applying our
conclusions to these cases can lead to significant modifications of
the results obtained in previous works \cite{amm,r}.

\section{Conclusion}
We have examined several issues related to the processes of Dirac
monopole-antimonopole production in high-energy collisions
such as $e^+e^-$ annihilation.
Perturbative calculations for such processes are known to be
inherently ambiguous due to the arbitrariness of direction
of the monopole string; this requires use of some prescription
to obtain physical results.
We argue that different prescriptions lead to  drastically
different physical results which suggests that at present we
do not have an entirely satisfactory procedure for the 
elimination of string arbitrariness (this problem is quite
separate from the problems caused by the large coupling
constant). We then analyze the consequencies of discrete
symmetries (P and C) for the monopole production processes and for
the monopole-antimonopole states. The P and C selection rules 
for the monopole-antimonopole states turn out to be different
from those for the ordinary fermion-antifermion or boson-antiboson
 systems.
In particular, the spin 1/2 monopole and antimonopole should have
{\em the same} helicities if they are produced through
one-photon annihilation of an electron and positron.
In the case of  {\em spinless} monopoles
CP symmetry {\em absolutely forbids} the monopole-antimonopole
production through the one-photon  annihilation
of an electron and positron.  
Single-photon $e^+e^- \rightarrow g^+g^-$ amplitude has been a key
input in calculating the contribution of virtual $g^+g^-$ pairs to
various physical processes such as the decay $Z\rightarrow 3\gamma$
and the anomalous magnetic moment of the electron. Applying our
conclusions to these cases can lead to significant modifications of
the results obtained in previous works.
This work was supported in part by the Australian Research Council.

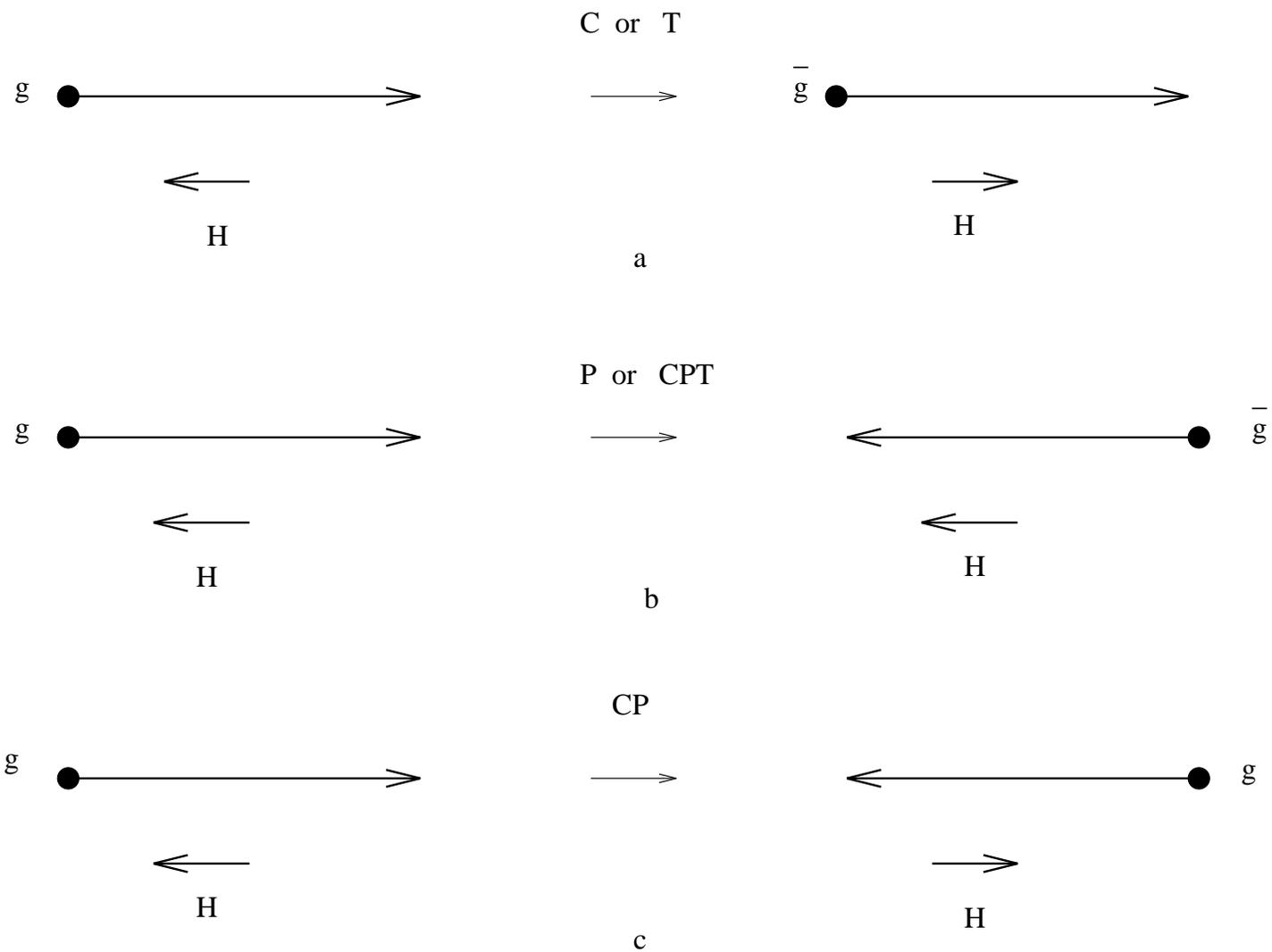
\begin{figure}
\caption{Action of the discrete symmetries on the 
semi-infinite string}
\label{fig3}
\end{figure}

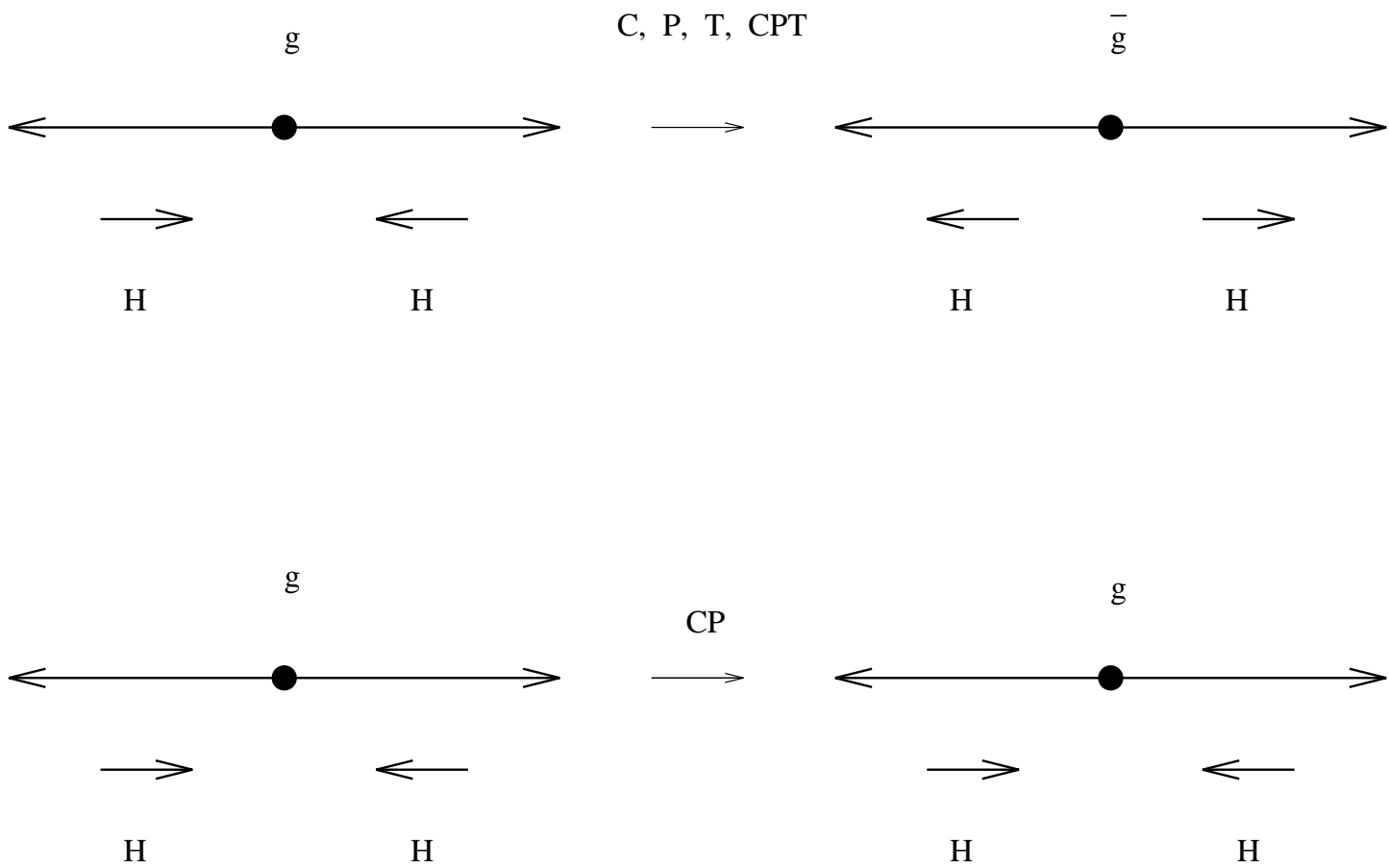
\begin{figure}
\caption{Action of the discrete symmetries on the infinite string}
\label{fig4}
\end{figure}

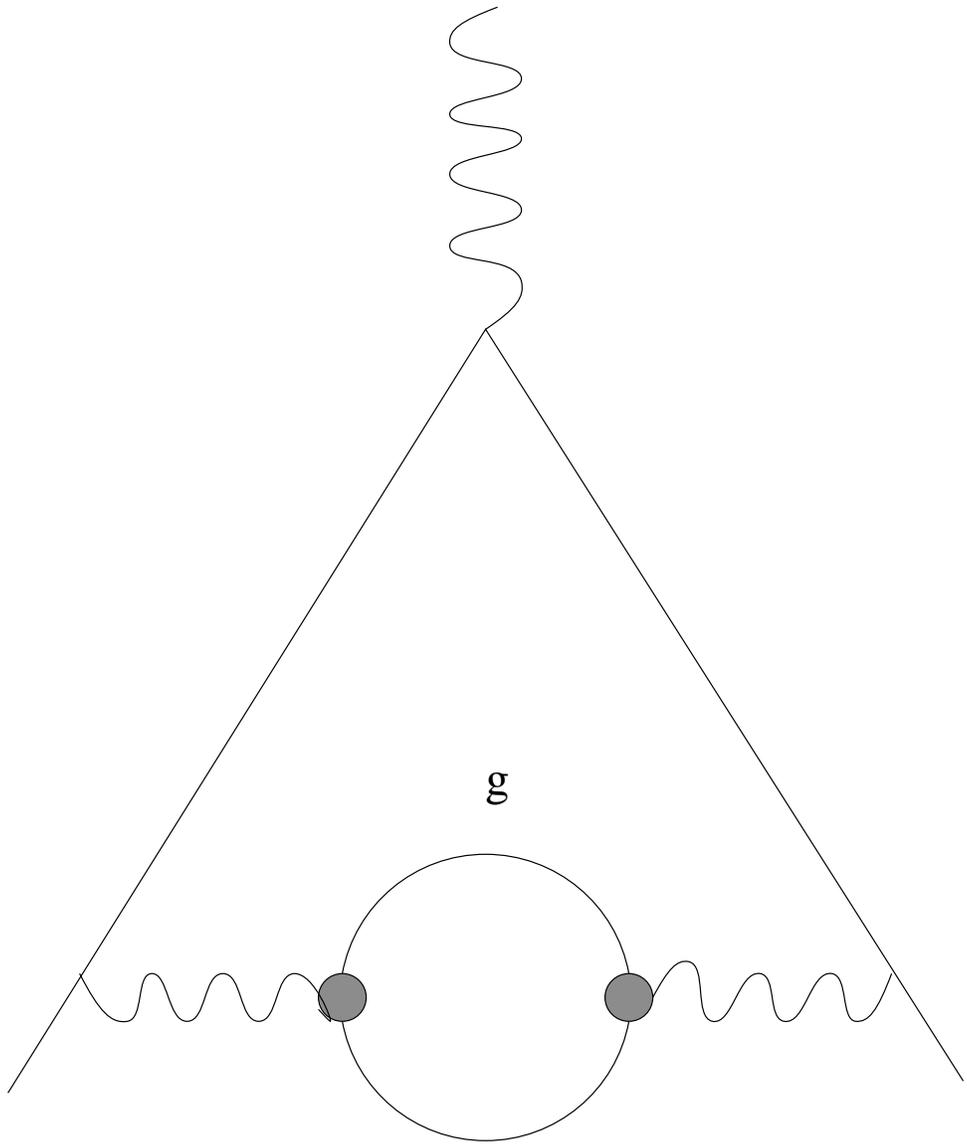
\begin{figure}
\caption{Contribution of virtual monopole-antimonopole pair to the 
anomalous
magnetic moment of the electron}
\label{fig5}
\end{figure}

\end{document}